\documentclass[twocolumn]{revtex4}
 \usepackage{verbatim}
\usepackage{graphicx}
\usepackage{float}
\usepackage{amssymb}
\usepackage[utf8]{inputenc}
\usepackage[english,french]{babel}
\usepackage[intlimits]{amsmath}
\usepackage{amsfonts}
\usepackage{bm}%
\usepackage{hyperref}

 \newcommand{\abs}[1]{\left\vert#1\right\vert}

\begin{document}
\title{Conformal invariance of loop ensembles under Kardar-Parisi-Zhang dynamics}
\author{Xiangyu Cao}
\affiliation{Laboratoire de Physique Théorique et Modèles Statistiques(UMR 8626), Université Paris Sud 11, CNRS, France}
\author{Alberto Rosso}
\affiliation{Laboratoire de Physique Théorique et Modèles Statistiques(UMR 8626), Université Paris Sud 11, CNRS, France}
\author{Raoul Santachiara}
\affiliation{Laboratoire de Physique Théorique et Modèles Statistiques(UMR 8626), Université Paris Sud 11, CNRS, France}
\begin{abstract}
We study scaling properties of the honeycomb fully packed loop ensemble associated with a lozenge tiling model of rough surface, when the latter is driven out of equilibrium by Kardar-Parisi-Zhang (KPZ) type dynamics. We show numerically that conformal invariance and signatures of critical percolation appear in the stationary KPZ state. In terms of the two-component Coulomb gas description of the Edwards-Wilkinson stationary state, our finding is understood as the invariance of one component under the effect of the non-linear KPZ term. On the other hand, we show a breaking of conformal invariance when the level lines of the other component are considered.
\end{abstract}
\date{\today}
\maketitle

\section{Introduction} Non-equilibrium stationary states in driven systems often display scale invariance, and an important question is whether it can be extended to conformal invariance. In particular, on the 2-d plane, the existence of conformal invariance, which is infinite-dimensional, would have powerful implications on our theoretical understanding of non-equilibrium states of $(2+1)$-d systems. Traces of conformal invariance are rare to find, and the answer to this question is unclear from a theoretical viewpoint.

One remarkable positive result was obtained by Bernard \textit{et al} \cite{bernard2006turbulence} in the context of 2-d incompressible Navier-Stokes turbulence. Strong numerical evidence suggested that zero-vorticity lines behave as cluster frontiers of critical percolation. The unexpected appearance of percolation in this context has raised great interest, yet still awaits a thorough understanding. In particular, does the presence of percolation teach us something on the turbulent state? Is there a link between it and other contexts where critical percolation signatures occur, \textit{e.g.}, nodal domains of wave functions in quantum chaos \cite{bogomolny2002percolation,bogomolny2007perc} or kinetic Ising ferromagnet \cite{olejarz2012fate}?

Recently, Saberi \textit{et. al.} \cite{saberi2008kpz} applied the idea (put forward initially by Kondev and Henley \cite{kondev1995loop}) of studying loop ensembles to the $(2+1)$-d Kardar-Parisi-Zhang(KPZ) equation \cite{kardar1986dynamic},
\begin{equation}\label{eq_kpz}
\partial_t h =  \vec{\nabla}^2 h + \frac{\lambda}{2} \abs{\vec{\nabla} h}^2 + \eta.
\end{equation} 
This is a fundamental model of irreversible surface growth, and it also describes the turbulence without pressure \cite{kardar1986dynamic, polyakov1995turbulence}, among other phenomena (for a recent review, see \cite{halpin2015kpz}). Saberi \textit{et. al.} observed numerically that level lines of stationary KPZ surface behave like self-avoiding walks, and so enjoy conformal invariance. This claim remains controversial \cite{hosseinabadi2013universality} and worrisome caveats \cite{saberi2010kpz} have not been resolved. 

Nevertheless, in this Letter we will show that, on KPZ rough surface models one can define a different loop ensemble which is conformal invariant and displays properties of critical percolation. In particular, it passes stringent tests involving $3$-point correlations, whereas KPZ level lines fail them. Moreover, we can understand our finding in terms of field theory. 

\section{Model and numerical implementation}\begin{figure} \center
\includegraphics[scale=0.18]{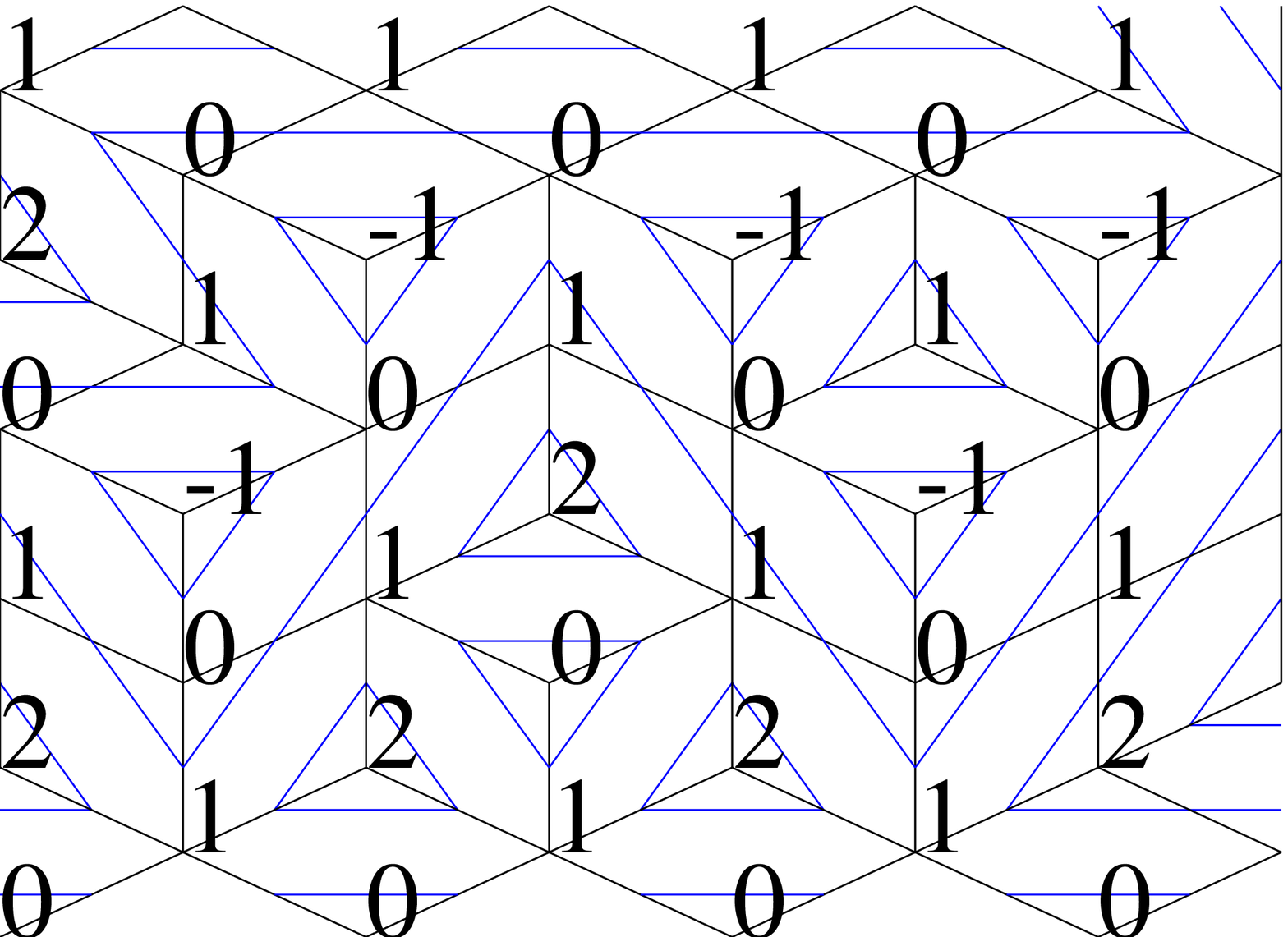} 
\text{ }\includegraphics[scale=0.18]{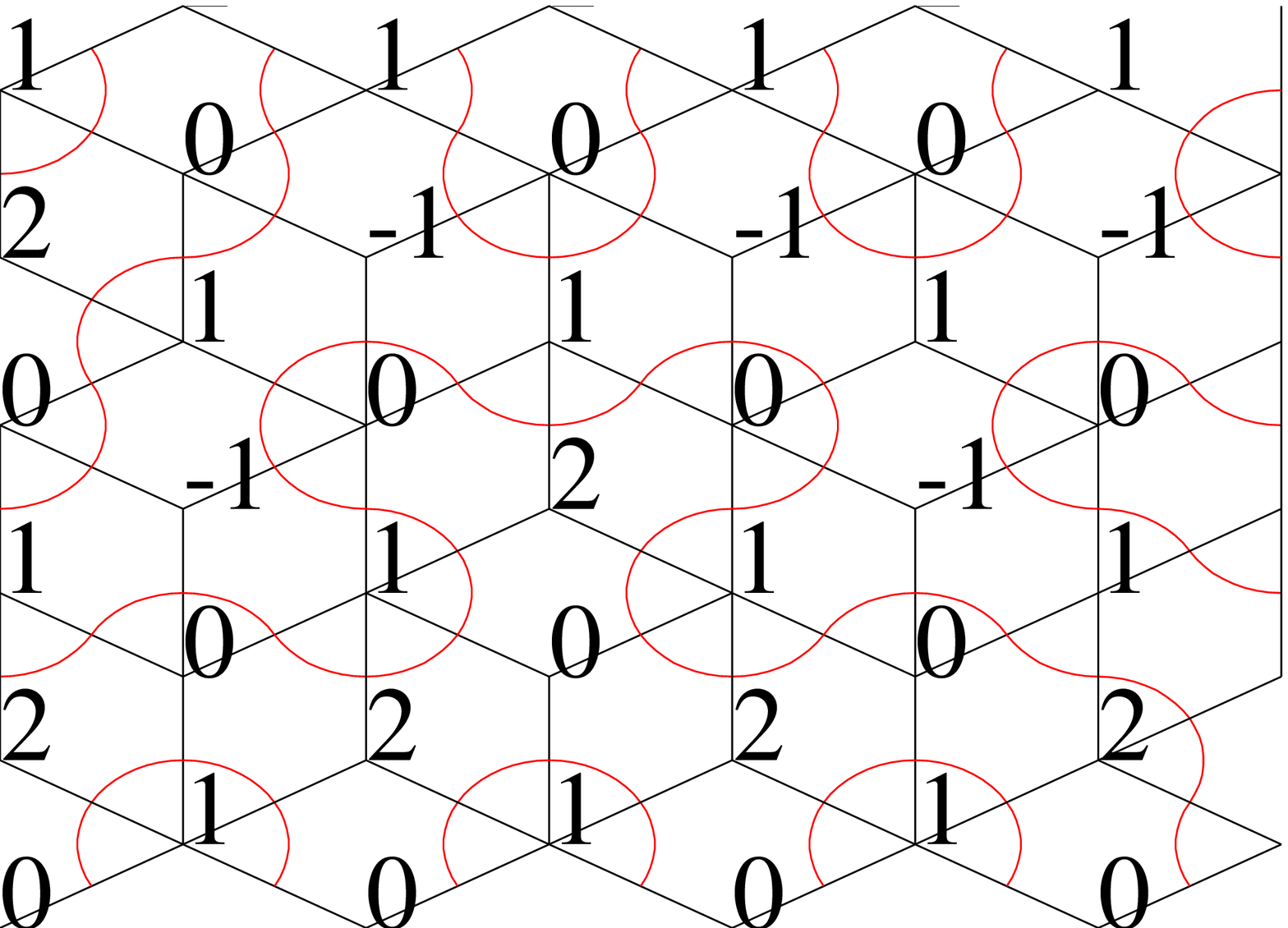} 
\caption{\textit{Left}: A local snapshot of the unit cube/lozenge rough surface model, with the surface height $h$ value on each vertex and level lines. \textit{Right}: The fully-packed loops constructed on top of the configuration. \label{fig_stat_t1}}
\end{figure}
We will focus on the model whose configurations are tiling of unit cubes of the cubic lattice (or, in a $2d$ viewpoint, lozenge tiling of the plane). As shown in Fig. \ref{fig_stat_t1} left, a height function $h$ is naturally defined on it. By linking middle points of each lozenge in this way \includegraphics[scale=1.5]{dessin.4} (and similarly for the other two lozenge kinds, as in Fig. \ref{fig_stat_t1} right), one establishes a bijection between lozenge tiling configurations and honeycomb-lattice fully packed loop configurations. Olejarz \textit{et. al.} \cite{olejarz2012growth} proposed the following stochastic unit cube deposition process: for each time interval $\mathrm{d}t$, an eligible move as shown below is chosen randomly:
$\includegraphics[scale=1.1]{dessin.2} \rightarrow \includegraphics[scale=1.1]{dessin.1}. $ 
This dynamics is known to be in the KPZ universality class \cite{halpin2012kpz}: The  coarse-grained limit of $h$ satisfies Eq. \ref{eq_kpz} in the continuum. Symmetrising the above dynamics, we obtain a growth model described by the Edwards-Wilkinson(EW) equation,\textit{i.e.}, KPZ equation (\ref{eq_kpz}) with $\lambda = 0$:
$\includegraphics[scale=1.1]{dessin.2} \rightleftharpoons  \includegraphics[scale=1.1]{dessin.1}.$ 

\begin{figure*}
\center
\includegraphics[scale=0.9]{dessin.13}
\includegraphics[scale=0.27]{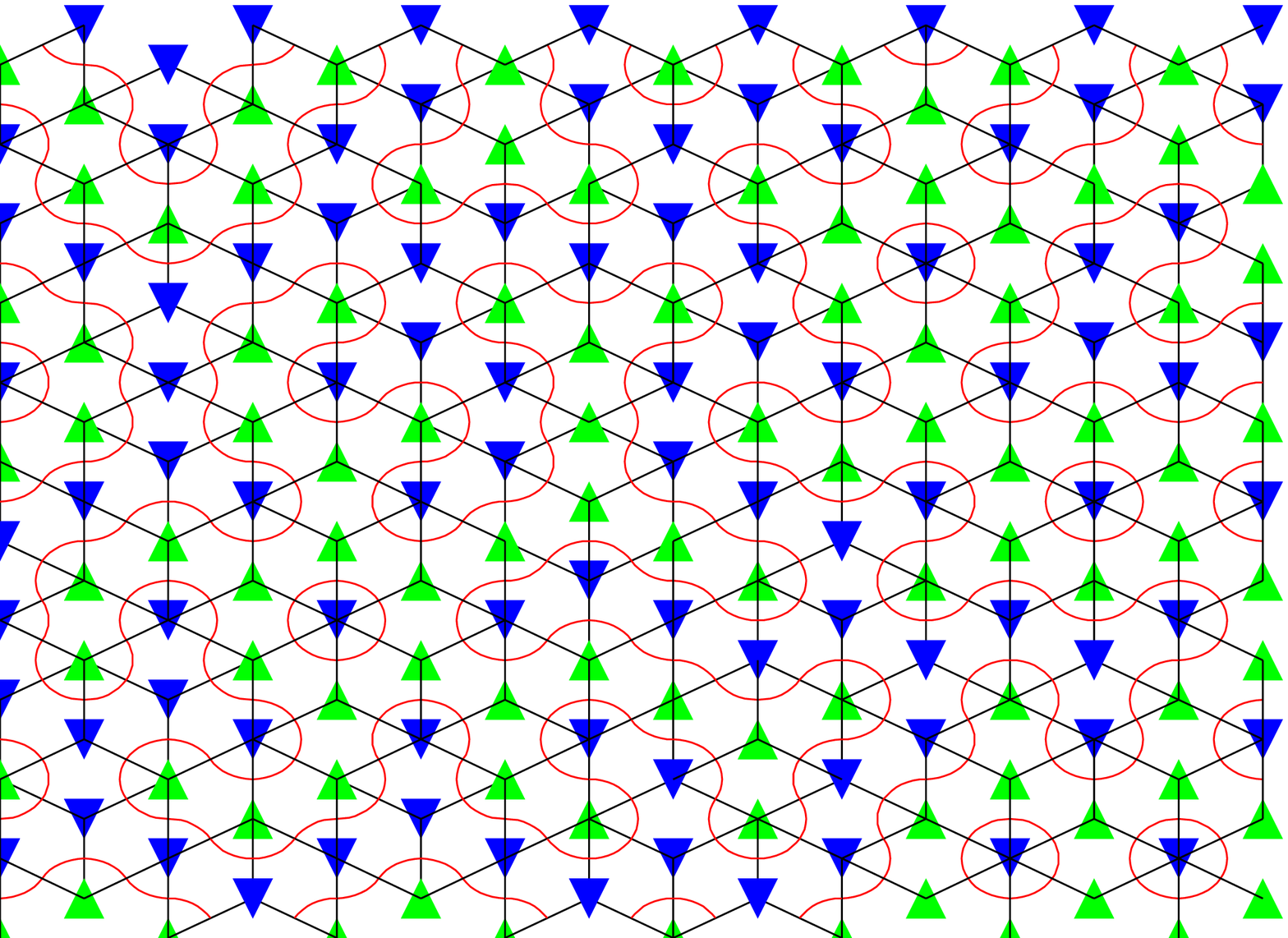} \textit{           }
\includegraphics[scale=0.27]{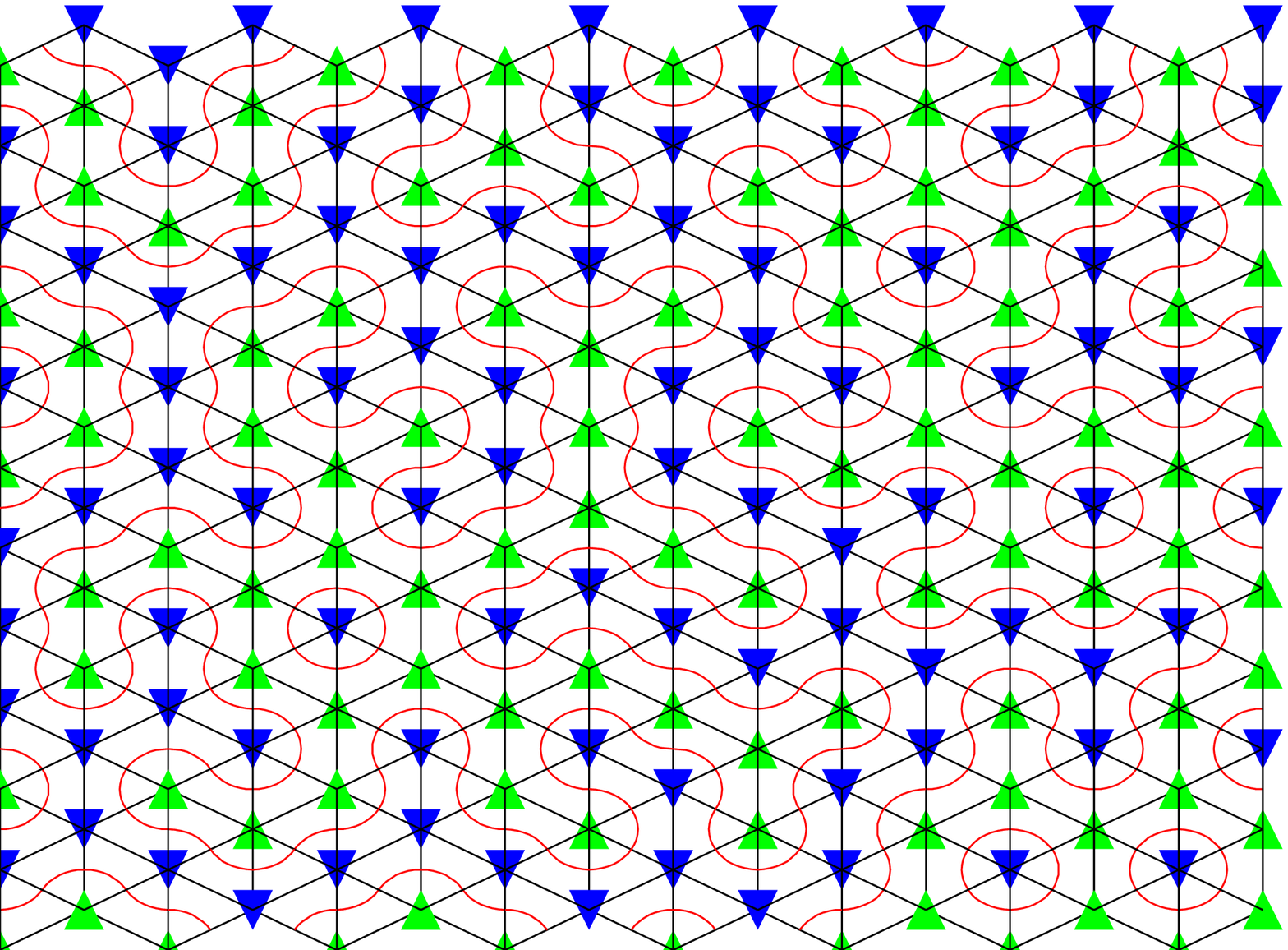}
\caption{\textit{Left}: Generators of the trigular lattice $e_1, e_2$ and that of a sublattice $E_1, E_2$. \textit{Middle/right}: A maximally anti-ferromagnetic spin configuration and its corresponding lozenge tiling. }\label{fig_spin}
\end{figure*}

It is interesting to represent lozenge tiling configurations in terms of Ising spin configurations on the triangular lattice. In fact, see Fig. \ref{fig_spin} middle and right, given such a spin configuration which is maximally anti-ferromagnetic (\textit{i.e.}, for each elemental triangle, the spins on its $3$ vertices are not the same), a lozenge tiling can be constructed by removing (from the triangular lattice) all the lattice edges connecting equal spins. The fully packed loops are then identical to the spin interfaces separating clusters of same spins. Therefore, standard methods for constructing spin clusters and boundaries can be applied to both lozenge tiling models and triangular-lattice site percolation. (In terms of the height $h$, the Ising spin is obtained by  $\sigma := \exp(\mathbf{i}\pi h)$, \textit{i.e.}, $\pm1$ according to parity of $h$.)
\begin{figure*}
\center
\includegraphics[scale=0.27]{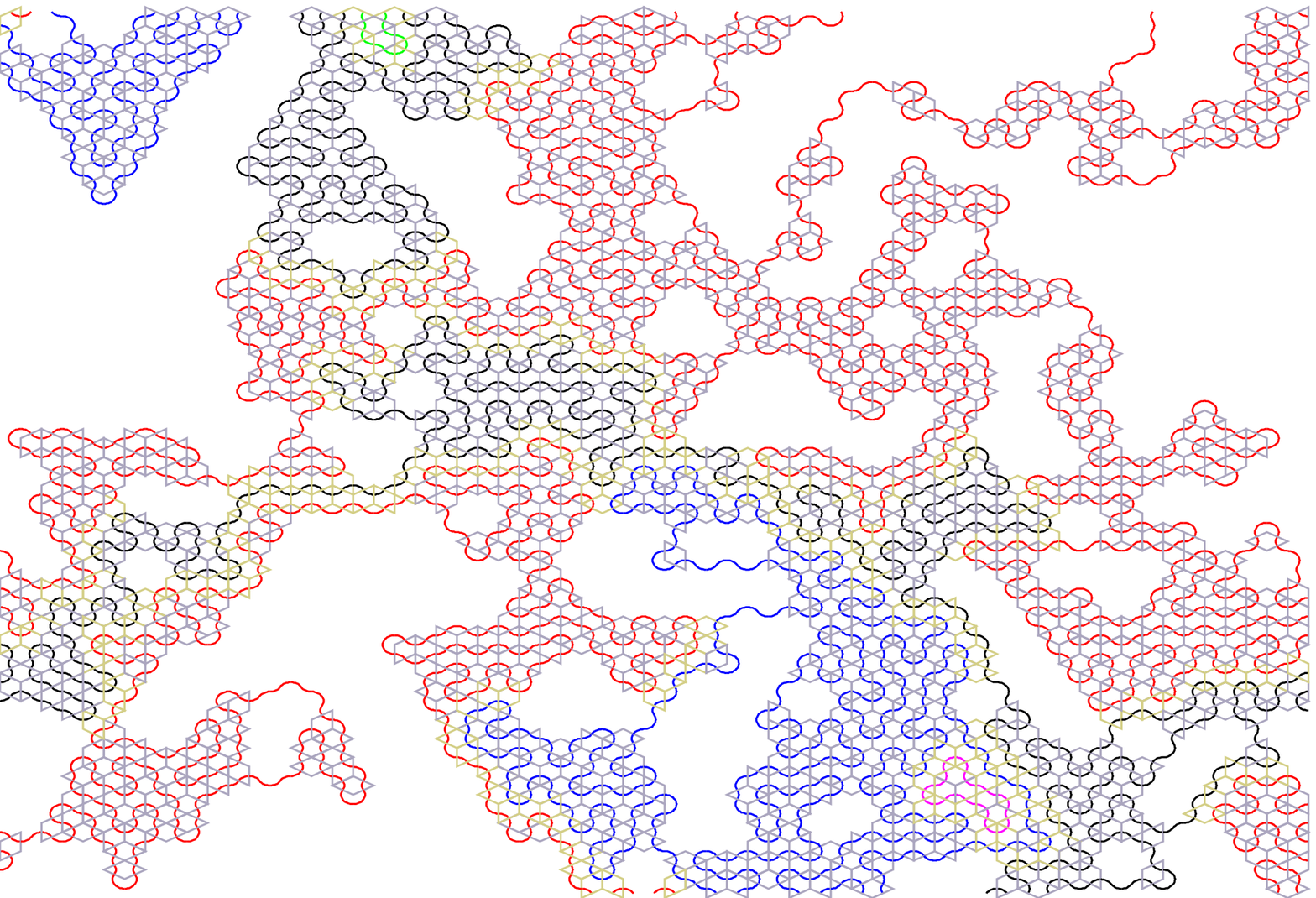}
\includegraphics[scale=0.25]{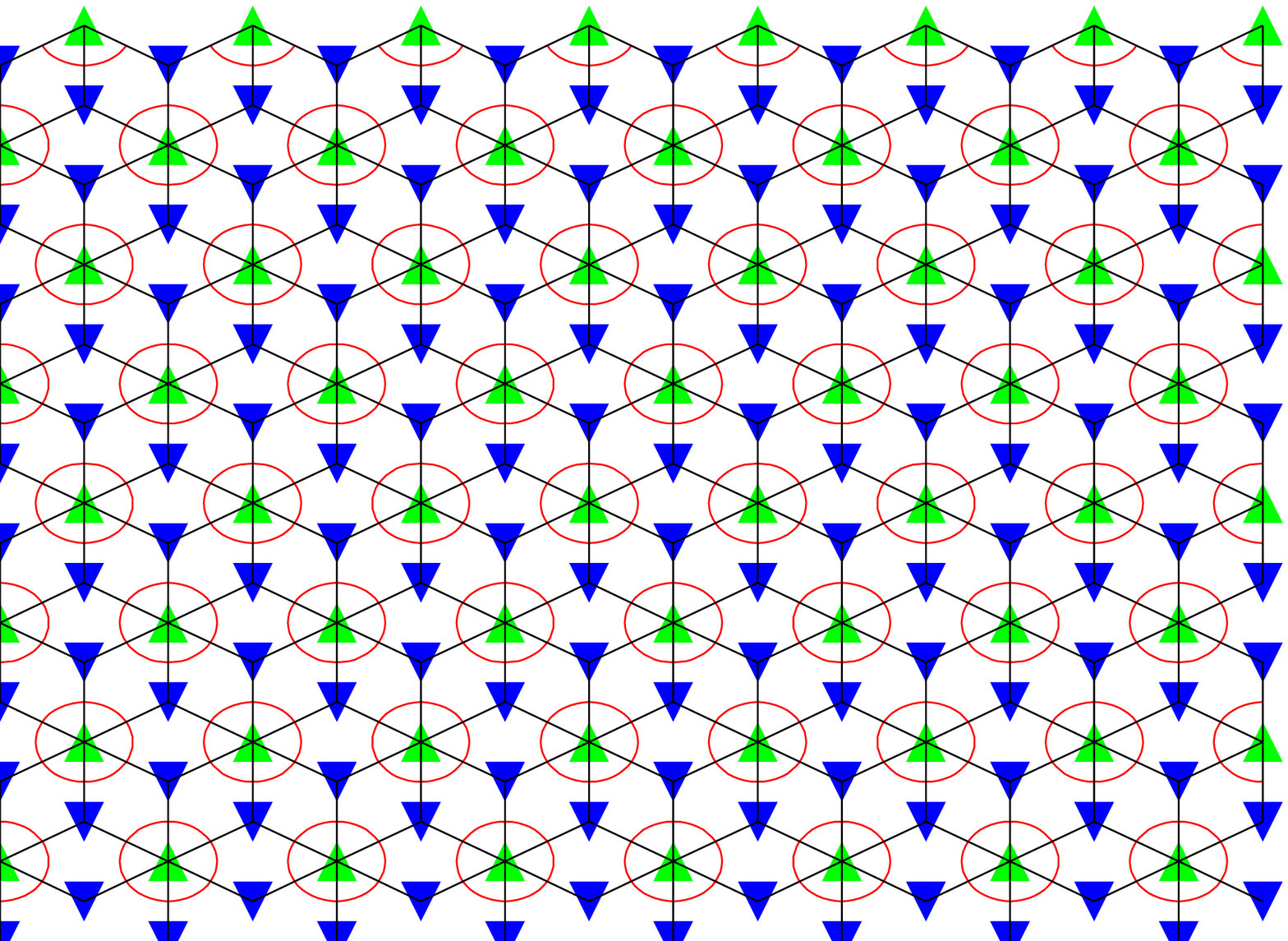}
\includegraphics[scale=0.27]{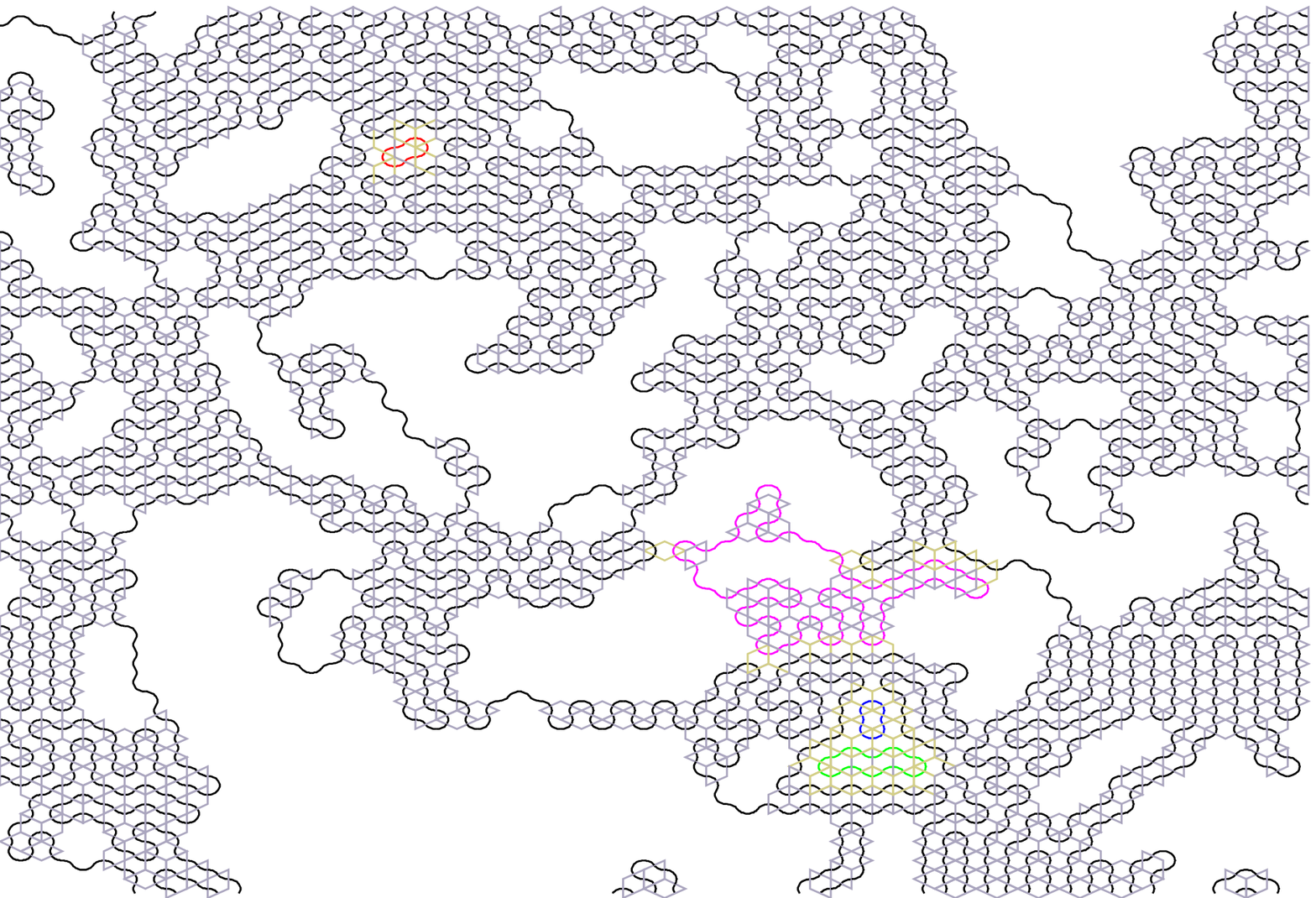}
\caption{\textit{Left}: Snapshot of some full packed loops in the stationary regime of EW.  Middle: The flat configuration.  \textit{Right}: Snapshot of some full packed loops in the stationary regime of KPZ. The snapshots are taken in system of size $L = 64$; distinct loops are drawn with different colors.}\label{fig_flat}
\end{figure*}

With this representation, the dynamics EW and KPZ are single-flip dynamics of Ising spins. In each trial, a site is picked at random and its spin $\sigma$ is compared to that of its $6$ nearest neighbours, $\sigma_0,\dots,\sigma_5$ (in circular order, with $\sigma_0$ pointing up). For EW, the central spin is flipped if $\sigma = \sigma_0 = \sigma_2 = \sigma_4$ \textit{or} if $\sigma = \sigma_1 = \sigma_3 = \sigma_5$ (this corresponds to a Metropolis Monte Carlo dynamics); for KPZ, the central spin is flipped only if $\sigma = \sigma_1 = \sigma_3 = \sigma_5$ (this is a non-equilibrium dynamics). It is not hard to see that in terms of lozenge tiling, the dynamics just described is equivalent to the respective ones defined above. The number of lozenges of each type are preserved by the dynamics, and we choose to stay in the sector with equal number of lozenges of each type. This is assured by starting from the flat initial condition depicted in Fig. \ref{fig_flat} middle. We let the system attain the stationary regime before constructing loop ensembles; snapshots of typical long loops are shown in Fig. \ref{fig_flat} left(EW) and right(KPZ).

Some care is needed to ensure the compatibility with the toroidal periodic boundary condition (p.b.c). Indeed, see Fig. \ref{fig_spin} left, let $e_1$ and $e_2$ be basis vectors of the triangular lattice edges forming $120$ degrees(in terms of Euclidean inner products $(e_1,e_1) =(e_2,e_2) = 1, (e_1,e_2) = -1/2$), then the triangular lattice is partitioned into three sub-lattices generated by $E_1 = e_1 + 2e_2 $ and $E_2 = e_1 - e_2$, and the flat configuration is the one with up spin on one sub-lattice and down on the other two (see Fig. \ref{fig_flat} middle). We fix our p.b.c for a system of size $L$ by identifying $ LE_1 \equiv LE_2 \equiv 0$, preserving the tri-partition and thus being compatible with the flat configuration. 
 
Data presented below are obtained on systems of size up to $L = 2048$. 
For each measure, $1e3 \sim 1e4$ samples are generated, extending over a time scale corresponding to $1e4 \sim 1e5$ elemental moves \textit{per lattice site}. 

\subsection{Fully packed loops and level lines} The stationary state of EW dynamics has the same weight for all tiling configurations and  coincides therefore with the well-known fully-packed loop model (which we shall call the EW FPL henceforth). Many results were obtained by Bethe Ansatz, \cite{batchelor1994exact} and Coulomb gas approaches \cite{kondev1996fpl}. The EW FPL loop ensemble is conformal invariant and strictly related to critical percolation. In particular, the loops have the same fractal dimension $D_{\text{f}} = 7/4$ (we define this notion below) as percolation frontiers. On the other hand, at the KPZ stationary state, the tiling configurations have different weights and no exact result is known. We will refer to the corresponding loop ensemble as KPZ FPL. We address the question whether the KPZ non-linear term affect the universality of EW FPL. For instance will the KPZ loop fractal dimension be different from the EW one? 

Let us recall that this is the case for \textit{level lines} of $h$, which are obtained by connecting lozenge middle points in the other way \includegraphics[scale=1.5]{dessin.3}, see Fig. \ref{fig_stat_t1} left. In the EW case, these lines are described in the continuous limit by the level lines of Gaussian free field and have therefore fractal dimension $D_{\text{f}}^{\text{lev.}}=3/2$ \cite{schramm2009contour}. In the KPZ case, the fractal dimension of the same lines was estimated numerically to be $D_{\text{f}}^{\text{lev.}} \approx 1.33$ \cite{saberi2008kpz}. Here we show that the critical percolation nature of the fully packed loop ensemble persists under the KPZ dynamics.
 
\section{Scaling exponents} 
\begin{figure}
 \includegraphics[scale=.42]{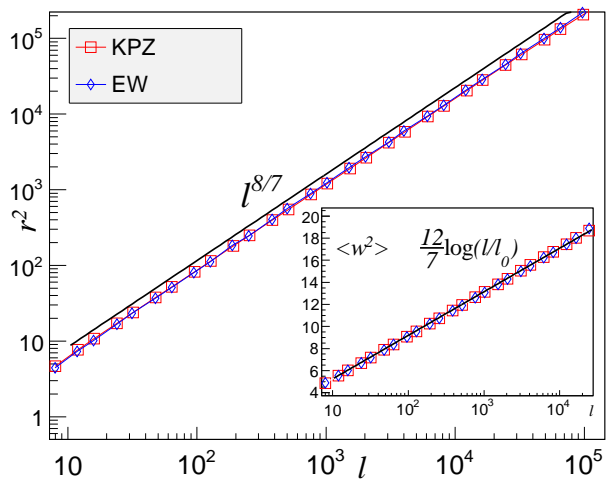}
 \caption{\label{fig_r2l}\textit{Main}: The mean square distance between extremities $\langle r^2 \rangle$($y$-axis) as function of loop segment length $l$ ($x$-axis). \textit{Inset}: Semi-log plot of the winding angle variance $\langle w^2 \rangle$ as function of $l$. Black lines are theoretical predictions (\ref{eq_r2}), (\ref{eq_w2}) with $D_{\text{f}} = 7/4$.}
 \end{figure}

The loop fractal dimension is related to a scaling relation between $r$, the distance between two extremities of a loop segment and $l$ its length:
\begin{equation}\label{eq_r2}
 \langle r^2 \rangle \sim l^{2/D_{\text{f}}}.
\end{equation}
This scaling behaviour holds when $a \ll l \ll L$, where $a$ is the lattice spacing and $L$ is the loop total length. In practice, we average over loop segments of length $ l < 0.1 L $ for $L > 5e3$. The result, shown in Fig. \ref{fig_r2l} main, suggests convincingly that fully-packed loops have $D_{\text{f}} = 7/4$ for both EW and KPZ stationary state. 

With the same loop segments at disposal, we also measure \textit{winding angle }variance. For a length-parametrised curve segment $r_s (s \in [0,l])$ of length $l$, let $\theta_s$ be a continuous function satisfying $\mathrm{d}r/\mathrm{d}s = (\cos \theta_s, \sin \theta_s)$, \textit{i.e.}, its tangent direction angle, then its (end-to-end) winding angle is defined as $w := \theta_l - \theta_0.$ It is predicted that(\cite{wieland2003winding})
\begin{equation}\label{eq_w2}
 \left\langle (\theta_l - \theta_0)^2 \right\rangle = \frac{4D_{\text{f}}}{(D_{\text{f}} - 1)} \log(l/l_0)
\end{equation}
(where $l_0$ is a lattice-dependent constant), holds for \textit{conformal invariant} curves of fractal dimension $D_{\text{f}}$. As shown in Fig. \ref{fig_r2l} inset, the above formula, with $D_{\text{f}} = 7/4$, agrees perfectly with KPZ and EW FPL. This strongly supports conformal invariance in the KPZ FPL ensemble. 
 
\begin{figure}
\center \includegraphics[scale=.42]{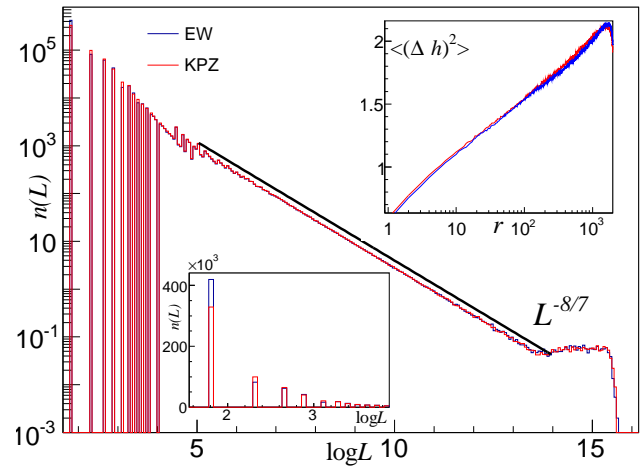}
\caption{\label{fig_tau} \textit{Main}: The distribution of loop length: $n(L)$, the number of loops with length $\in [L, L(1 + \epsilon))$($y$-axis,$\epsilon = 0.06$) as a function of $\log L$($x$-axis). Black Line is the theoretical prediction \ref{eq_nl} with $\tau = 15/7$. \textit{Up inset}: Roughness of the hidden surface $h_2$: The mean squared difference of the hidden surface $\Delta h_2 :=  h_2(\mathbf{r}) - h_2(\mathbf{0})$ in function of spatial separation $r =\abs{\mathbf{r}}$. \textit{Down inset}: Zoom-in of the microscopic regime where EW and KPZ FPLs are different.}
\end{figure}

We consider next the loop length distribution, expected to have a power law behaviour
\begin{equation}\label{eq_nl}
\langle n(L) \rangle \sim L^{-\tau + 1}.
\end{equation}  
The above scaling should be valid in an interval $a \ll L \ll L_{\max}$, $L_{\max}$ being the system size cut-off. The definition of the exponent $\tau$ is the same as in \cite{kondev1995loop}. For critical percolation frontiers, it is known exactly that $\tau = 15/7$. In Fig. \ref{fig_tau} main we superpose $\langle n(L) \rangle$ of KPZ and EW surface. The two distributions are indistinguishable in the scaling regime, where the power law is again in agreement with the percolation prediction.
Using general scaling arguments, Kondev and Henley \cite{kondev1995loop} derived a relation between the roughness exponent of a scaling invariant surface, and the exponents $D_{\text{f}}$ and $\tau$ associated to its level lines
\begin{equation}\label{eq_kondev1}
D_{\text{f}}(\tau - 1) = 2 - \alpha.
\end{equation} 
Setting $D_{\text{f}} = 7/4, \tau = 15 / 7$ one gets $\alpha = 0$, which is true for any equilibrium critical loop model. But the fact that the same happens to KPZ is \textit{a priori} unexpected. Note that this is not in contradiction with the non-zero roughness of the KPZ surface, as we recall that surface level lines form a different ensemble from FPL.
One can give a meaning to the exponent $\alpha$ in Eq. (\ref{eq_kondev1}) by constructing a surface whose level lines are the FPLs. In practice, we assign a height increment crossing each loop by choosing randomly one of the values $\pm 1$. The obtained height function \footnote{Strictly speaking, \textit{complex} weights(instead of real probabilities) should be assigned in this construction. This technicality will not affect the log-roughness observed(but will affect the pre-factor).}
 we denote $h_2$. This field is the second component of the vector-valued height function $(h_1, h_2)$ in the Coulomb gas \cite{kondev1996fpl} approach to the FPL model. The other component $h_1$ is the lozenge height function $h$ (see Fig. \ref{fig_stat_t1} left) introduced from the beginning.

From a CFT viewpoint, Dotsenko \textit{et al} \cite{dotsenko2001classification} remarked that EW FPL can be described by the following effective action:
\begin{equation}\label{eq_cft_fpl} \mathcal{S} [h_1, h_2] = \mathcal{S}_1[h_1] + \mathcal{S}_2[h_2],
\end{equation} where the $h_1$ and $h_2$ degrees of freedom are decoupled (the decoupling was first found numerically by \cite{blote1994fully}). The action $\mathcal{S}_2[h_2]$ describes the critical percolation (dense $\mathcal{O}(n = 1)$ loop model, \cite{nienhuis1982exact}), while $\mathcal{S}_1[h_1]$ is a free-boson CFT with central charge $c = 1$. The KPZ stationary state eludes any field-theoretical description, being far from equilibrium. Nevertheless, we put forward the hypothesis that \textit{correlation functions in the $h_2$ sector of the EW FPL are not affected by the KPZ dynamics, while those involving the field $h_1$ are.}
The results shown above supports this scenario. In particular, the field $h_2$ has logarithmic roughness for KPZ as well as for EW. This is also confirmed by explicit numerical measure of mean squared height difference in function of spatial separation Fig.  \ref{fig_tau} up inset. In the following, we provide further evidence of the above hypothesis by looking at $3$-point correlation functions.

\section{$3$-point structure constants and conformal invariance} Until now, our hypothesis is corroborated by observables that can be related to $2$-point functions. Yet truly profound predictions of CFT begin with $3$-point functions. As illustrated in Table \ref{ill_3p}, for $k > 1$, we consider the probability $P_l(x_1, x_2, \dots, x_k)$ (\textit{l}oop $k$-point function) that $x_1, x_2, \dots,$ and $x_k$ lie on the same loop, and $P_c(x_1, x_2, \dots, x_k)$ (\textit{c}luster $k$-point function) the probability that, for any couple of points $x_i, x_j$, there exists a curve connecting them and not crossing any loop. 
\begin{table}
\center 
\begin{tabular}{c|c|c}
$P_l(x_1, x_2, x_3) = 1$ & $ P_c(x_1, x_2, x_3) = 1$ & $P_c(x_1, x_2, x_3) = 0$  \\
\hline
\includegraphics[scale=1]{dessin.10} & \includegraphics[scale=1]{dessin.11} & \includegraphics[scale=1]{dessin.12}
\end{tabular}
\caption{\label{ill_3p} From left to right: a configuration contributing to $P_l(x_1, x_2, x_3)$, a configuration contributing to $P_c(x_1, x_2, x_3)$, and a configuration \textit{not} contributing to $P_c(x_1, x_2, x_3)$} 
\end{table}

When $k = 2$, the CFT predicts
$ P_{o}(0,x) \sim \abs{x}^{-2 x_{o}}, o = c,l,$
with $x_l = 1/4$ and $x_c = 5/48$ \cite{nienhuis1982exact}. For $k = 3$,
We consider the \textit{$3$-point ratio}
\begin{equation}\label{eq_3p2p}
C_o = \frac{P_{o}(x,y,z)}{\sqrt{P_{o}(x,y)P_{o}(x,y)P_{o}(x,y)}}, o=c,l.
\end{equation}
Global conformal invariance implies that $C_o$ is a \textit{constant}, universal, 
which defines the CFT. 

For critical percolation, the exact value of $C_c$ has been predicted by Delfino and Viti \cite{delfino2011three} to be given by the Liouville CFT \cite{ribault2015liouville}. This conjecture was numerically confirmed \cite{delfino2011three,picco2013connectivities}. The exact value of $C_l$ is not known (see, however, \cite{estienne2015correlation}), and to our knowledge numerical values are not published. \footnote{Using transfer-matrix techniques, Y. Ikhlef and J. L. Jacobsen have obtained numerical estimates of $C_l$ which coincides with ours by $3$ digits. (Private communication)}

\begin{figure}
 \center \includegraphics[scale=.435]{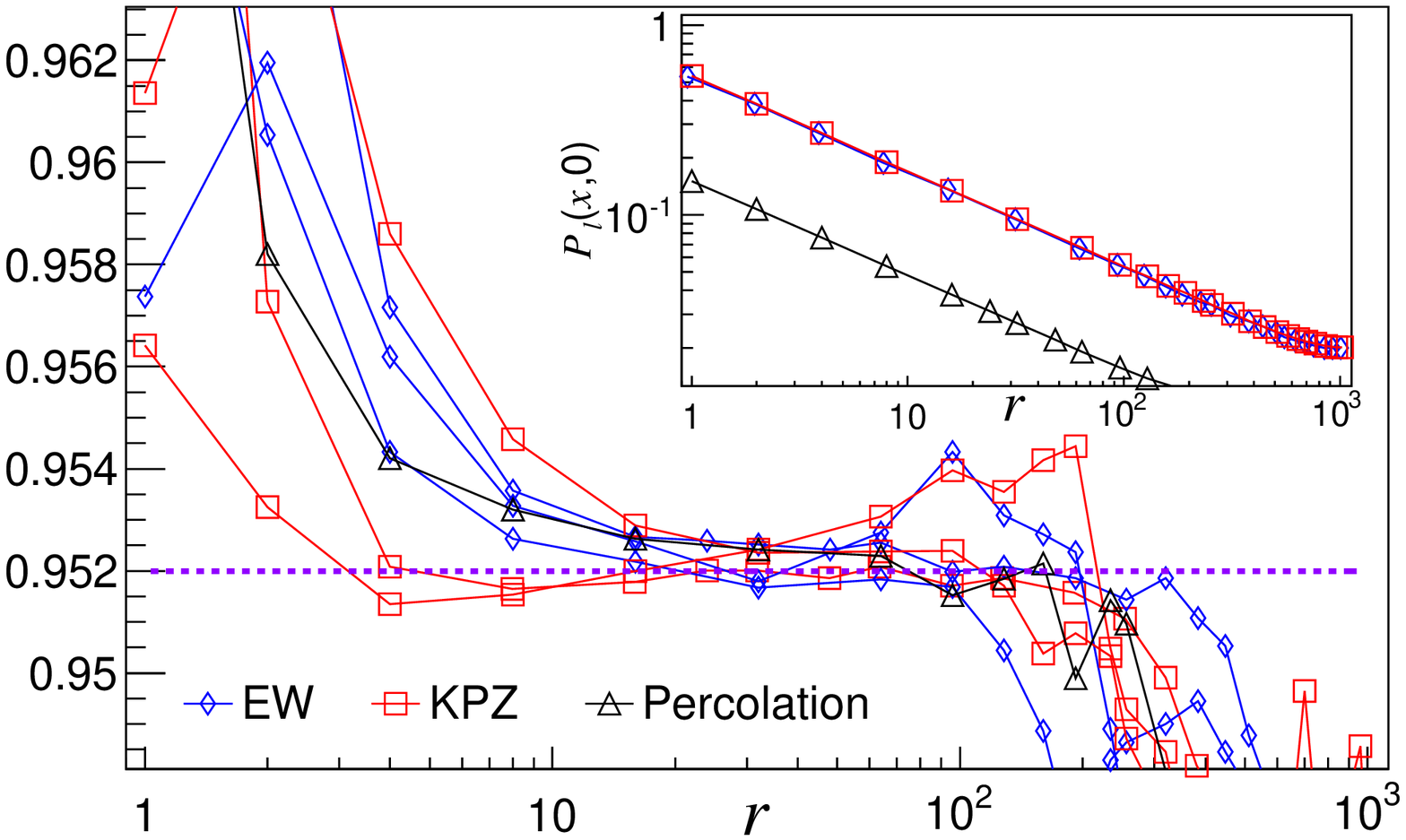}\vspace*{0.1cm}
 \includegraphics[scale=.435]{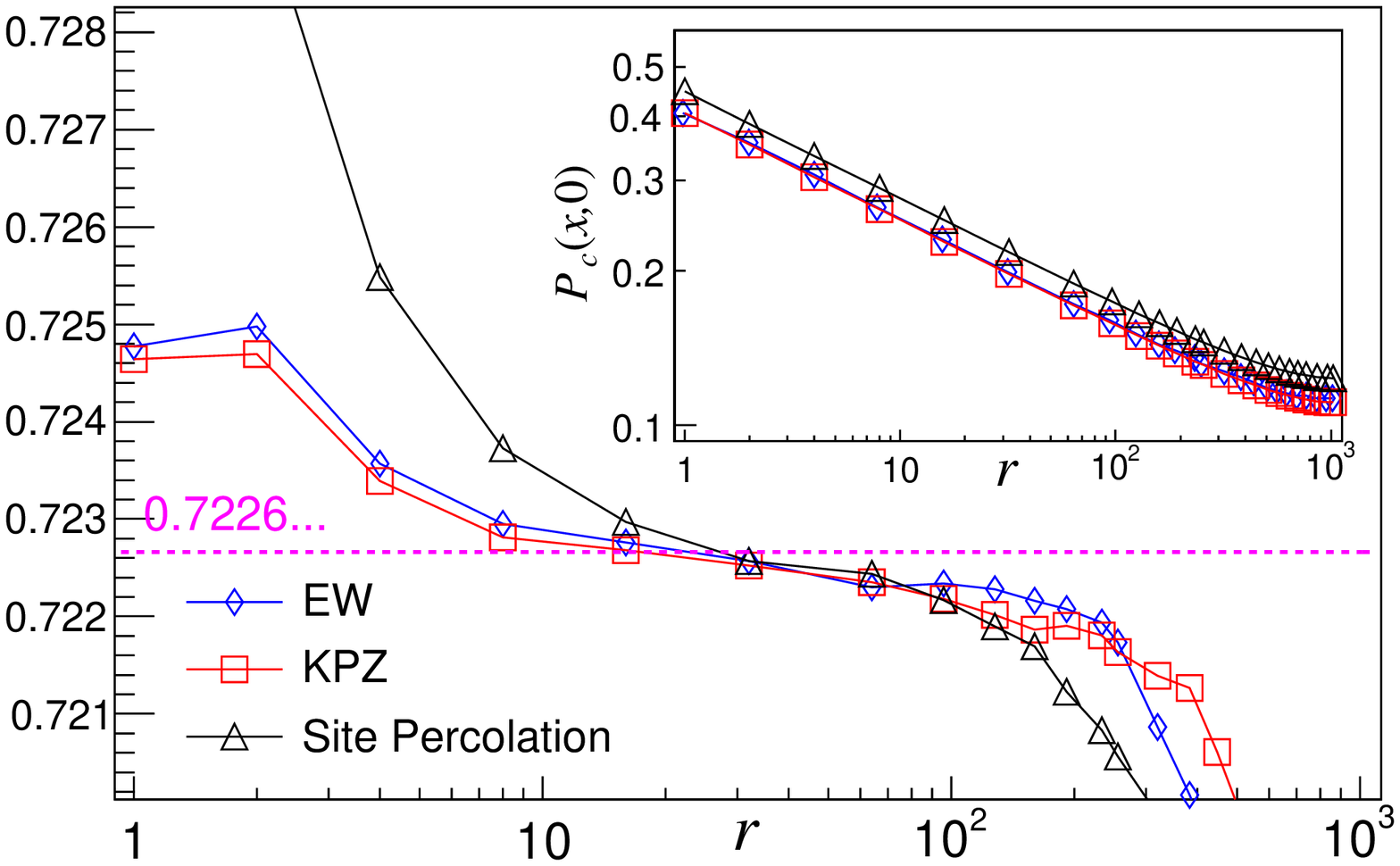}
 \caption{\label{fig_cl3p}
\textit{Main frames}: $3$-point ratios where the points $x,y$ and $z$ form a equilateral triangle of side $r$. \textit{Up}: Loop $3$-point ratio estimate $C_l=0.952(3)$. Measures using different microscopic definitions of point are superposed. \textit{Down}:  Cluster $3$-point ratios compared to prediction $C_c \approx 0.722\dots$. \textit{Insets}: the corresponding $2$-point function in function of distance $r$.}
 \end{figure}

We determine the ratios $C_c$ and $C_l$ for EW and KPZ FPLs. The $3$-point ratios are calculated for $x,y$ and $z$ forming equilateral triangles of varying radius; different geometries have been checked for the independence of $C_o$. 
The results are shown in the main plots of Fig. \ref{fig_cl3p}, and compared to those obtained from critical site percolation frontiers on the triangular lattice. 
The numerics support strongly the following: 
\begin{itemize}
\item[•] $2$-point functions $P_{(l,c)}$ of EW and KPZ FPLs agree with the percolation exponents $x_l = 1/4, x_c = 5/48$.
\item[•] It exists a scaling regime where the $3$-point ratios are constant for EW and KPZ FPLs. 
\item[•] The value of $C_c$, which is the same for EW and KPZ FPLs, and the numerical result is in excellent agreement with Delfino-Viti's prediction for the critical percolation: 
\begin{equation} C_c = 0.722\dots 
\end{equation}
\item[•] The loop structure constant $C_l$ is the same for critical percolation, EW and KPZ FPLs, its numerical value is estimated to
\begin{equation} C_l = 0.952(3).
 \end{equation}
We further collapsed this estimate against an independent measure on critical bond percolation on the square lattice.
\end{itemize}
The above results are in agreement with our hypothesis that the conformal invariance in the $h_2$ sector is not broken by KPZ dynamics. Nonetheless, we remark that the EW-KPZ coincidence of the observables considered occurs not only in universal exponents, but often also in non-universal amplitudes. This appears far from trivial and begs an elementary explanation. 

\section{$h_1$ sector observables} 
\begin{figure*}
\center 
\includegraphics[scale=0.35]{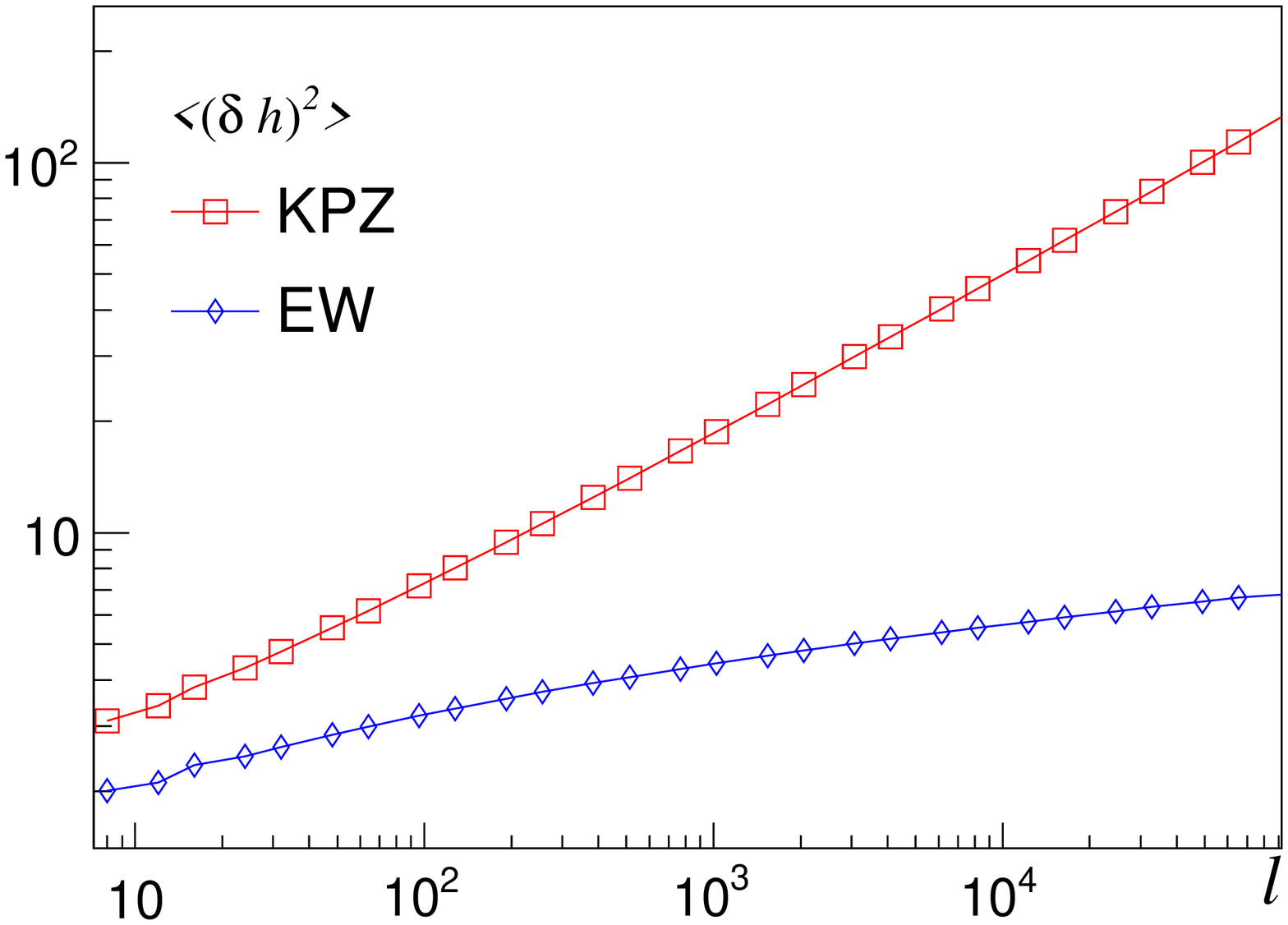} 
\includegraphics[scale=0.37]{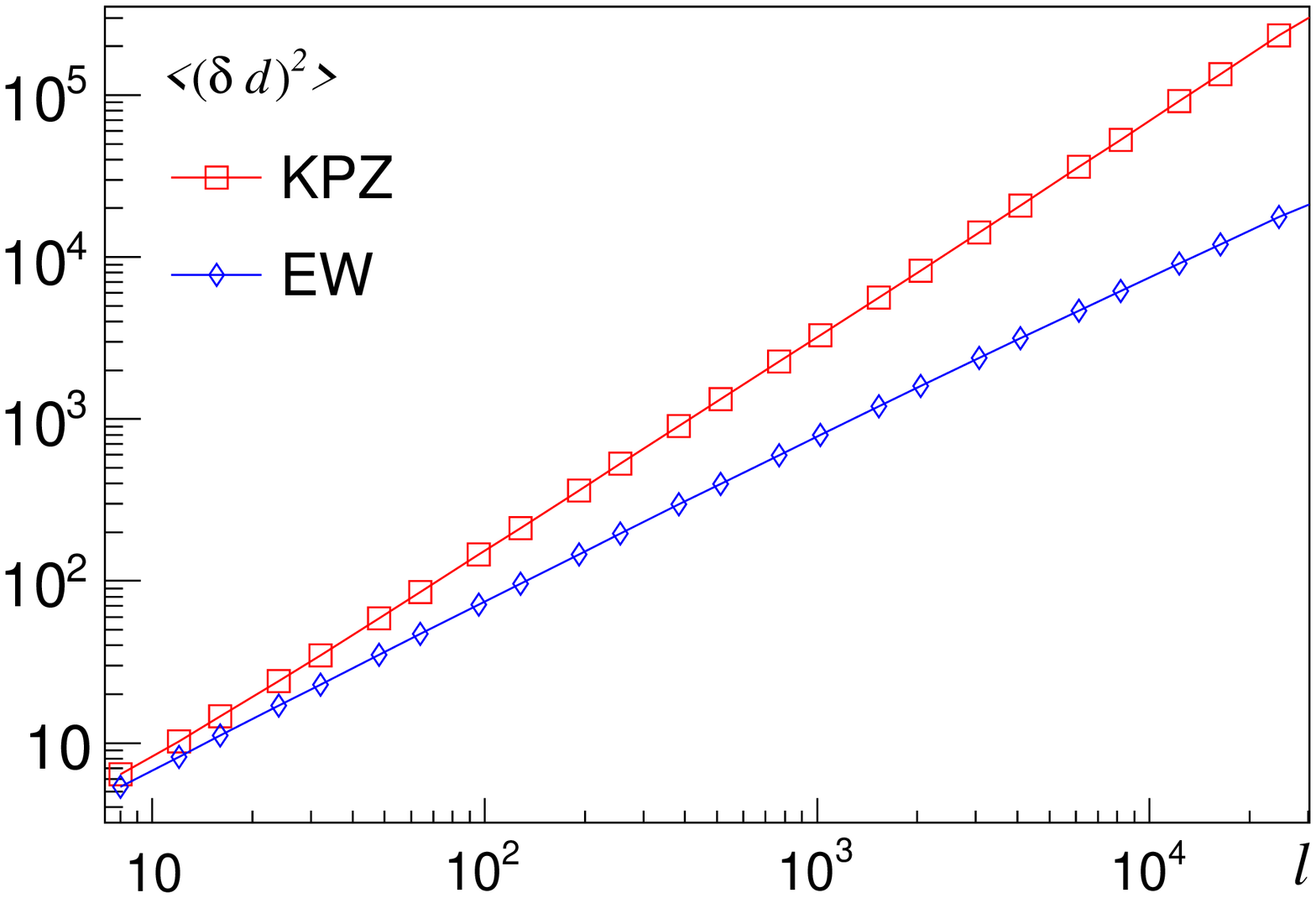} 
\caption{\textit{Left}: the observable $(\delta h)^2$ as function of loop segment length $l$. \textit{Right}: the observable $(\delta d)^2$ as function of loop segment length $l$.\label{fig_lrugo}}
\end{figure*}

Contrary to what can be suggested by the results of the previous section, we emphasize that the EW and KPZ FPLs are \textit{not identical}. It indeed suffices to encode some information on $h_1$ in the FPLs to see distinct behaviours for EW and KPZ. Here we show how to construct such observables of \textit{individual} loops. For this, consider a loop segment of length $l$. It can be described by a sequence of left/right turns of $60$ degrees, $ \mathrm{d} \theta_s := \theta_s - \theta_{s-1} = \pm \frac{\pi}{3}, s = 1, \dots, l.$ (The fully packed loops live in fact on the hexagonal lattice dual to the triangular lattice.) 
Recall that the winding angle, defined as $\sum_{s=1}^l  \mathrm{d}\theta_s$ associated to the loop segments cannot distinguish KPZ from EW for $l$ large. However, the observable $\delta h({l}) := \frac{3}{\pi} \sum_{s=1}^l (-1)^s \theta_s$ does so, see Fig. \ref{fig_lrugo} left. The variance $\langle (\delta h)^2 \rangle$ has logarithmic growth for EW and power-law for KPZ. Indeed, it can be shown that $ |\delta h(l)| = |h_1(r_0) - h_1(r_l)|$ where $r_0$ and $r_l$ are the extremity positions of the loop segment. Thus, $\delta h(l)$ encodes the roughness of $h_1$ (logarithmic for EW, positive power law for KPZ) into fully packed loops, explaining the observed distinct scaling behaviour. Note that the definition of $\delta h$ differs from the winding angle $w$ just by an alternative sign. To produce a second example, we consider the extremity displacement $r(l) := \sum_{s=1}^l \mathrm{d} r_s$, where $\mathrm{d} r_s = (\cos \theta_s, \sin \theta_s)$, which cannot distinguish KPZ and EW at long distance. On the contrary, the quantity $\delta d(l) := \sum_{s=1}^l (-1)^s \mathrm{d} r_s$ grows with distinct power laws for KPZ and EW fully packed loops (Fig. \ref{fig_lrugo} right). Again, one can realise that this measure encodes information from $h_1$, whose roughness is also reflected in correlation of lozenge orientation. Denoting $n_0,n_1$ and $n_2$ the number of lozenges of each orientation passed by the loop segment, it is not hard to see that $(\delta d)^2= 
\abs{n_0 + n_1 e^{i 2/3\pi} + n_2e^{-i 2/3\pi}}^2$ and measures the deviation of $n_i$'s from the expected average $l/3$. 

 \begin{figure}
 \center \includegraphics[scale=.40]{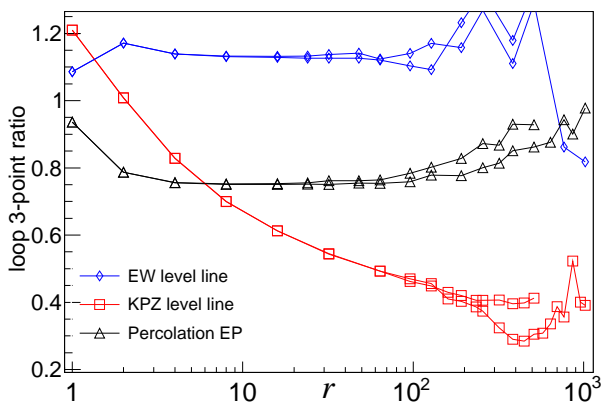}
 \caption{\label{fig_saberi} Loop $3$-point ratios of EW and KPZ \textit{level line} ensemble as well as the external perimeters ensemble of critical percolation. Different curves of the same type indicate the error.}
 \end{figure} 
We finally consider the $3$-point ratio for the \textit{level lines} (of height $h_1$), which is of more physical interest, see Fig. \ref{fig_stat_t1} left. Saberi \textit{et. al.} conjectured that the KPZ level lines are in the universality class of self-avoiding walks. Here we compare loop $3$-point ratio of KPZ level lines with that of \textit{external} perimeters of critical site percolation clusters, since they are believed to be identical to self-avoiding walks \cite{duplantier1999harmonic,aizenman1999path}. As we show in Fig. \ref{fig_saberi}, percolation external perimeters display a well-defined loop structure constant, whereas KPZ level lines do not. Finally, we observe that the loop $3$-point ratio for EW level lines attains also a constant. 
This result, taken together with another incompatibility (noticed in \cite{saberi2010kpz}) with hyper-scaling relations in \cite{kondev1995loop} and estimates of KPZ roughness exponents \cite{kelling2011extremely}, casts doubt on the conformal invariance of KPZ level lines and their conjectured identification to self-avoiding walks. 

\section{Conclusion} In this work, we have studied KPZ stationary state compared to the EW one on a natural discrete model. We have shown that, in terms of the two-component surface description, whereas $h_1$ undergoes dramatic change under the KPZ dynamics, the ``hidden'' component $h_2$ remains invariant. The fully packed loop ensemble associated with $h_2$ enjoys conformal invariance, and has critical properties indistinguishable from that of critical percolation. On the other hand, conformal invariance appears broken for level lines of $h_1$. The decoupling of $h_2$ from $h_1$ in KPZ occurs exactly in the scaling regime, even if at the microscopic level, there is no \textit{a priori} reason. We are left to wonder what implications, if any, our observations can have on field-theory of non-equilibrium stationary states in general, and on the genuine KPZ phenomenology in particular. 

\acknowledgments
We are pleased to acknowledge K. Mallick for showing us the model studied in this work.
Moreover, we thank E. Bogomolny, Y. Ikhlef, P. L. Krapivsky, P. Le Doussal, M. Picco and J. Viti for useful discussions. 

\bibliography{biblio.bib}
\end{document}